\documentclass[amstex,aps,preprint,nofootinbib]{revtex4}%
\usepackage{graphicx}
\usepackage{color}
\usepackage{bm}
\usepackage{braket}
\usepackage{longtable}
\usepackage{amsmath}
\usepackage{amsfonts}
\usepackage{amssymb}
\usepackage{float}
\usepackage{epstopdf}%

\begin{document}
\title{Nonlocality and local causality in the Schr\"{o}dinger Equation with
time-dependent boundary conditions}
\author{A. Matzkin}
\affiliation{Laboratoire de Physique Th\'{e}orique et Mod\'{e}lisation (CNRS Unit\'{e}
8089), Universit\'{e} de Cergy-Pontoise, 95302 Cergy-Pontoise cedex, France}
\author{S. V. Mousavi }
\affiliation{Department of Physics, University of Qom, Ghadir Blvd., Qom 371614-6611, Iran}
\author{M. Waegell}
\affiliation{Institute for Quantum Studies, Chapman University, Orange, CA 92866, USA}

\begin{abstract}
We investigate the nonlocal dynamics of a single particle placed in an
infinite well with moving walls. It is shown that in this situation, the
Schr\"{o}dinger equation (SE) violates local causality by causing
instantaneous changes in the probability current everywhere inside the well.
This violation is formalized by designing a gedanken faster-than-light
communication device which uses an ensemble of long narrow cavities and weak
measurements to resolve the weak value of the momentum far away from the
movable wall. Our system is free from the usual features causing nonphysical
violations of local causality when using the (nonrelativistic) SE, such as
instantaneous changes in potentials or states involving arbitraily high
energies or velocities. We explore in detail several possible artifacts that
could account for the failure of the SE to respect local causality for systems
involving time-dependent boundary conditions.

\end{abstract}
\maketitle

\section{Introduction}

Nonlocality is the hallmark of quantum mechanics. It is generally taken for
granted that nonlocality requires two or more particles, along the lines of
the early paper by Einstein Podolsky and Rosen \cite{EPR}, subsequently put
into a firm footing by Bell \cite{bell}. Although it has been suggested that a
single particle could in some instances exhibit nonlocality, such results have
been disputed. This is particularly the case of the two main candidates for
single particle nonlocality, the Aharonov-Bohm effect \cite{AharonovAB} and
the entanglement between spatial modes of a single photon (see \cite{lee2017}
and Refs. therein for previous works), discussed respectively in Refs.
\cite{vaidmanAB,kangAB} and \cite{bjork2001,vedral2007}.

The present work introduces a new ``candidate'' for single particle
nonlocality. It is based on the fact that the Schr\"{o}dinger equation solved
on a domain with moving boundaries gives rise to apparent violations of local
causality. It appears that time-dependent boundary conditions can potentially
induce a nonlocal change in a region located far from the location of the
moving boundary. Here we will examine the case of a particle in a box with
infinitely high but moving walls.\ We will see that for quantum states
extended all over the box, the moving walls generate instantaneously a current
density almost everywhere in the box. We will indicate how this effect could
be in principle tested, namely by making weak measurements of the particle
momentum in the central region of the box before light has the time to
propagate from the walls to that region. To this effect, a gedanken
faster-than-light communication device will be presented.

Let us state right away that we are not advocating the position that it is
possible to send a signal faster than the speed of light. Nevertheless, the
present problem is interesting because the non-relativistic Schr\"{o}dinger
equation fails to prevent superluminal signaling in a situation where
relativistic considerations do not seem to play a significant role. It is
indeed well-known that the Schr\"{o}dinger equation does not bound particle
velocities, nor does it constrain instantaneous changes in potentials, but we
will argue that in our system the nonlocal aspects do not rely on spurious
violations of special relativity allowed by a employing a nonrelativistic framework.

Note that the effect reported in this work is not due to a non-dynamical phase
term, such as a geometric phase (in which case we would have in the present
context a non-adiabatic, non cyclic geometric phase \cite{samuel,mukunda}).
There have been in the past claims that such non-dynamical phases in the same
type of system that we will be investigating in this work could be envisaged
as a specific form of ``hidden'' (i.e., non-signaling) non-locality
\cite{aharonov-tokyo,greenberger,aharonov-rohrlich}. We will see instead that
the non-local aspect in our candidate system is not based on the existence of
such phases.

We will start by revisiting the treatment of systems with time-dependent
boundary conditions of the form $\psi(x(t),t)=0$, where $\psi$ is the
wavefunction.\ Such systems are delicate to handle because from a formal point
of view a different Hilbert space needs to be defined for each time $t$, so
that a simple operation like taking the time derivative $\partial_{t}\psi$ is
not straightforward. We will introduce the system we will deal with -- a
particle in an expanding infinite well -- in the context of recent works
\cite{martino2013,art2015,A2018} involving time dependent boundary conditions
in Sec.\ 2.

Weak measurements were originally \cite{AAV} introduced to measure an
observable without significantly disturbing the system, allowing a subsequent
standard (projective) measurement of a different observable. The outcome,
known as a weak value, is not generally an eigenvalue (since the quantum state
of the system is barely modified and no projection takes place) but still
gives some information on the weakly measured observable, provided enough
statistics are gathered by repeating the experience a certain number of
times.\ In particular, it was shown \cite{leavens} that the weak value of the
momentum is directly related to the current density. We will recall these
facts in Sec. \ref{sec3} where we will present our main results concerning the
instantaneous response of the current density to a change in the boundary conditions.

We will then proceed (Sec.\ 4) to analyze and discuss this novel type of
nonlocality. The first issue we will address is no-signaling. No-signaling
stands as the major constraint permitting the \textquotedblleft peaceful
coexistence\textquotedblright\ \cite{shimony} of relativity and quantum
mechanics. At first sight it would appear that no-signaling is respected
here, since a single weak measurement does not convey any information, but the
situation is more involved, and a protocol that would allow us to test in
principle the possibility of signaling will be presented. Given that this
nonlocal effect appears to conflict with the no-signaling principle, we will
critically assess the origins of nonlocality, in search of possible artifacts.
We will then discuss the present results in the framework of the Bohmian
model, where nonlocality is a built-in feature claimed to hold for individual
events but is washed out at the statistical level. A summary and our
conclusions will be given in Sec. \ref{conclu}.

\section{A particle in an infinite well with moving walls \label{sec2}}

The particle in an infinite well with moving walls was widely investigated in
the context of quantum chaos (see e.g.  \cite{seba,physica,glasser,robnik}).
Another line of studies concerning this system involves the conjecture of
nonlocality induced by the moving wall on a localized state
\cite{greenberger,makowski1992,zou2000,yao2001,wang2008,mousavi2012,mousavi2014}%
, that was recently disproved \cite{A2018}.\textbf{\ }The Hamiltonian for a
particle of mass $m$ in an infinite well with the left wall fixed at $x=0$ and
the right wall moving according to the function $L(t)$ is given by%
\begin{align}
H  &  =\frac{P^{2}}{2m}+V\label{ham}\\
V(x)  &  =\left\{
\begin{array}
[c]{l}%
0\text{ \ for}\ \ 0\leq x\leq L(t)\\
+\infty\text{ \ otherwise.}%
\end{array}
\right.  . \label{vdef}%
\end{align}

The solutions of the Schr\"{o}dinger equation $i\hbar\partial_{t}%
\psi(x,t)=H\psi(x,t)$ must obey the boundary conditions $\psi(0,t)=\psi
(L(t),t)=0$. The instantaneous eigenstates of $H$,
\begin{equation}
\phi_{n}(x,t)=\sqrt{2/L(t)}\sin\left[  n\pi x/L(t)\right]  \label{eig}%
\end{equation}
verify $H\ket{\phi_{n}}=E_{n}(t)\ket{\phi_{n}}$ where $E_{n}(t)=n^{2}\hbar
^{2}\pi^{2}/2mL^{2}(t)$ are the instantaneous eigenvalues, but, due to the
time varying boundary conditions, the $\phi_{n}$ are \emph{not} solutions of
the Schr\"{o}dinger equation. To solve the Schr\"{o}dinger equation different
approaches have been proposed, like introducing a covariant time derivative
\cite{pershogin91}, implementing an ad-hoc change of variables
\cite{makowski91}, or relying on a time-dependent quantum canonical
transformation \cite{mosta1999,martino2013}. Here we follow the latter option,
as implemented in Ref. \cite{A2018}. However, rather than going through the
transformation to derive the solutions for the general case (this is done in
\cite{A2018}), we will choose from the beginning a specific function $L(t)$
for which analytic basis solutions of the Schr\"{o}dinger equation are known.
Indeed, for the linearly expanding case
\begin{equation}
L(t)=L_{0}+qt
\end{equation}
it can be checked by inspection \cite{makowski91} that%
\begin{equation}
\psi_{n}(x,t)=\sqrt{\frac{2}{L_{0}+qt}}\exp\left(  -\frac{i\pi^{2}\hbar
^{2}n^{2}t-iL_{0}m^{2}qx^{2}}{2\hbar mL_{0}\left(  L_{0}+qt\right)  }\right)
\sin\left(  \frac{n\pi x}{L_{0}+qt}\right)  \label{fund}%
\end{equation}
verifies the Schr\"{o}dinger equation and the boundary conditions
$\psi(0,t)=\psi(L(t),t)=0$. Here, $q>0$ represents the velocity of the
expanding wall.

The set of $\psi_{n}(x,t)$ (with $n$ a positive integer) form a set of
orthogonal basis functions useful to determine the time evolution of an
initial arbitrary quantum state. The simplest initial state would be to pick a
given $\psi_{n}(x,t=0);$ its evolution follows directly from Eq.
(\ref{fund}).\ From a physical standpoint, it would be more realistic to start
from the standard fixed wall eigenfunctions.\ A typical initial state woud
then be an eigenstate $\phi_{n}(x,t=0)$ [see Eq. (\ref{eig})] or a linear
combination thereof, say
\begin{equation}
\psi(x,t=0)=\sum_{n=1}^{\infty}c_{n}\phi_{n}(x,t=0) \label{stationis}%
\end{equation}
whose evolution is given by%
\begin{equation}
\psi(x,t)=\sum_{k,n}c_{n}\left\langle \psi_{k}(t=0)\right\vert \left.
\phi_{n}(t=0)\right\rangle \psi_{k}(x,t). \label{expan}%
\end{equation}
\ We may want to include additional refinements, like allowing for a
continuous transition from the fixed walls to the linear regime by setting
\begin{equation}
L(t)=L_{0}+qt(1-e^{-\gamma t}). \label{modL}%
\end{equation}
This requires numerical solutions. The numerical method that will be used here
is very similar to the one exposed in Ref. \cite{glasser}; it is based on
looking for numerical solutions $\zeta(x,t)$ by using expansions over the
instantaneous eigenstates of the form%
\begin{equation}
\zeta(x,t)=\sum_{k=1}^{\infty}a_{k}(t)\phi_{k}(x,t).
\end{equation}
The coefficients $a_{k}(t)$ are retrieved by solving a system (arising by
plugging $\zeta(x,t)$ in the Schr\"{o}dinger equation) of coupled differential equations.

\section{Current density and momentum weak values \label{sec3}}

\subsection{Current density evolution}

We first briefly look at the standard current density%
\begin{equation}
j\mathbf{=}\frac{1}{2m}\left(  \psi^{\ast}P\psi-\psi P\psi^{\ast}\right)  ,
\end{equation}
where $P$ is the momentum operator for the states in an expanding infinite
well. When the initial state is taken to be an eigenstate $\phi_{n}(x,0)$ of
the fixed walls well, given by Eq. (\ref{eig}) with all the $c_{k}$ vanishing
except for $c_{n}=1$, the current density is initially zero, but becomes
non-zero for $t>0$. Indeed, near the wall, the initial wavefunction is
nonzero, since $\phi_{n}(x\approx L_{0},t=0) \simeq n\left(  L_{0}-x\right) $,
and is substantially modified when the wall moves. By the arguments given in
\cite{A2018} (or simply by the continuity of the logarithmic derivative noting
that the potential remains unchanged except at $x=L(t)$) we then know that at
an infinitesimal time $t=\varepsilon$ we will have $\psi_{n}(x,\varepsilon
)-\psi_{n}(x,0)\neq0$ at any $x,$ although we expect this quantity to be large
near $x=L(\varepsilon)$ and smaller in the regions away from the moving wall.

When the initial state is taken to be a basis state $\psi_{n}$ given by Eq.
(\ref{fund}), the current density is immediately computed as%
\begin{equation}
j_{\psi_{n}}(x,t)=\frac{2qx\sin^{2}\frac{n\pi x}{L(t)}}{L(t)^{2}}.
\label{jfundn}%
\end{equation}
We see that $j_{\psi_{n}}(x,t)$ changes continuously both in the space and
time variables. The change in the current density at $x \ll L_{0} $ can easily
be computed.\ From
\begin{equation}
\Delta j(x)\equiv j(x,\varepsilon)-j(x,0), \label{supra0j}%
\end{equation}
we have
\begin{equation}
\Delta j_{\psi_{n}}(x)\underset{x\ll L_{0}}{\simeq}2\pi^{2}n^{2}qx^{3}\left(
\frac{1}{(L_{0}+q\varepsilon)^{4}}-\frac{1}{L_{0}^{4}}\right)  \underset
{\varepsilon\rightarrow0}{\simeq}-\frac{8\pi^{2}n^{2}q^{2}x^{3}}{L_{0}^{5}%
}\varepsilon. \label{supraj}%
\end{equation}
Let $t_{S}=\left(  L_{0}-x\right)  /c$ be the time it takes for a light signal
emitted at the wall to reach the point $x$ where the current density is
monitored ($c$ is the light velocity). Then there is a range of times
$\varepsilon$ such that $\varepsilon<t_{S}\ $and $\Delta j(x)\neq0$: the
current density is modified instantaneously by the wall's motion. The
significance of this instantaneous appearance of a current density will be
discussed in Sec. \ref{discussion}.\ We next examine how this current density
could in principle be experimentally tested.

\subsection{Weak measurements and the current density}

The underlying idea at the basis of the weak measurement (WM) framework
\cite{AAV} is to give an answer to the question:\textquotedblleft\textit{what
is the value of a property (represented by an observable $A$) of a quantum
system while it is evolving from an initial state }$\mathit{\left\vert
\psi(t_{i})\right\rangle }$\textit{\ to a final state $\left\vert b_{f}%
(t_{f})\right\rangle $?}\textquotedblright. This is done by coupling the
system observable $A$ to a dynamical variable of an external pointer, say
$\emph{Q}$ through an interaction Hamiltonian of the form
\begin{equation}
H_{int}=g(t)A\emph{Q}, \label{Hint}%
\end{equation}
where $g(t)$ is a smooth function nonzero during the interaction time.\ The
effective coupling constant $g\equiv\int g(t)dt$ is chosen to be
very small so that although the system and the external pointer
become entangled, the system state is minimally disturbed by $H_{int}$. A
standard measurement of another system observable, say $B$ can then be
undertaken.\ Assume that the eigenvalue $b_{f}$ corresponding to the
eigenstate $\left\vert b_{f}\right\rangle $ is obtained -- a step called
postselection.\ It can then be shown (\cite{AAV}; see e.g.  Sec.\ II of
\cite{Duprey2017} for a brief derivation) that the external pointer that was
coupled to $A$ has shifted by the quantity $gA^{w}$ where%
\begin{equation}
A^{w}=\frac{\left\langle b_{f}\right\vert A\left\vert \psi(t_{i})\right\rangle
}{\left\langle b_{f}\right\vert \left.  \psi(t_{i})\right\rangle } \label{wv}%
\end{equation}
is known as the \emph{weak value} of $A$ given the initial (preselected) state
$\left\vert \psi(t_{i})\right\rangle $ and the final (postselected) state
$\left\vert b_{f}\right\rangle $. \footnote{For simplicity we have disregarded
in Eq. (\ref{wv}) the evolution of the system between the initial preparation
time $t_{i}$, the mean interaction time $t_{w}$ and the postselection time
$t_{f};$ otherwise Eq. (\ref{wv}) should be replaced by
\begin{equation}
A^{w}=\frac{\left\langle b_{f}\right\vert U(t_{f},t_{w})AU(t_{w}%
,t_{i})\left\vert \psi(t_{i})\right\rangle }{\left\langle b_{f}\right\vert
U(t_{f},t_{i})\left\vert \psi(t_{i})\right\rangle }%
\end{equation}
(see e.g.  Sec.\ II of \cite{Duprey2017}).} Note that while $A^{w}$ is generally
a complex quantity, when one weakly measures observable $A$, the shift of the
external pointer is proportional to the real part of $A^{w}$. \footnote{The
imaginary part of $A^{w}$ is proportional to the shift of the momentum of the
pointer wavefunction --- for a pointer in position space.}

Let us now specialize Eq. (\ref{wv}) to a weak measurement of the momentum $P$
immediately followed by a standard measurement of the position, denoting the
outcome by $x$.\ The weak value is then given by $P^{w}=\frac{\left\langle
x\right\vert P\left\vert \psi\right\rangle }{\left\langle x\right\vert \left.
\psi\right\rangle }.$ It is easy to see that $P^{w}$ can be written as
\cite{leavens,wiseman,matzkin2012}
\begin{equation}
P^{w}=\frac{mj_{\psi}(x,t)}{\left\vert \psi(x,t)\right\vert ^{2}}-i\hbar
\frac{\partial_{x}\left(  \left\vert \psi(x,t)\right\vert ^{2}\right)
}{2\left\vert \psi(x,t)\right\vert ^{2}}. \label{pw}%
\end{equation}
Hence the real part of the momentum weak value is the hydrodynamic velocity
(well known from the Bohmian model, see Sec. \ref{bohm} below) $v(x,t)$ given
by%
\begin{equation}
v(x,t)\equiv\frac{j_{\psi}(x,t)}{\left\vert \psi(x,t)\right\vert ^{2}}%
=\frac{\operatorname{Re}{P}^{w}}{m}. \label{bv}%
\end{equation}
Our statement made above on the superluminal change in the current density
\ following the walls' motion has now been couched in terms of an
experimentally measurable quantity, the momentum weak value.

\begin{figure}[ptb]
\includegraphics*[angle=-90,width=16cm]{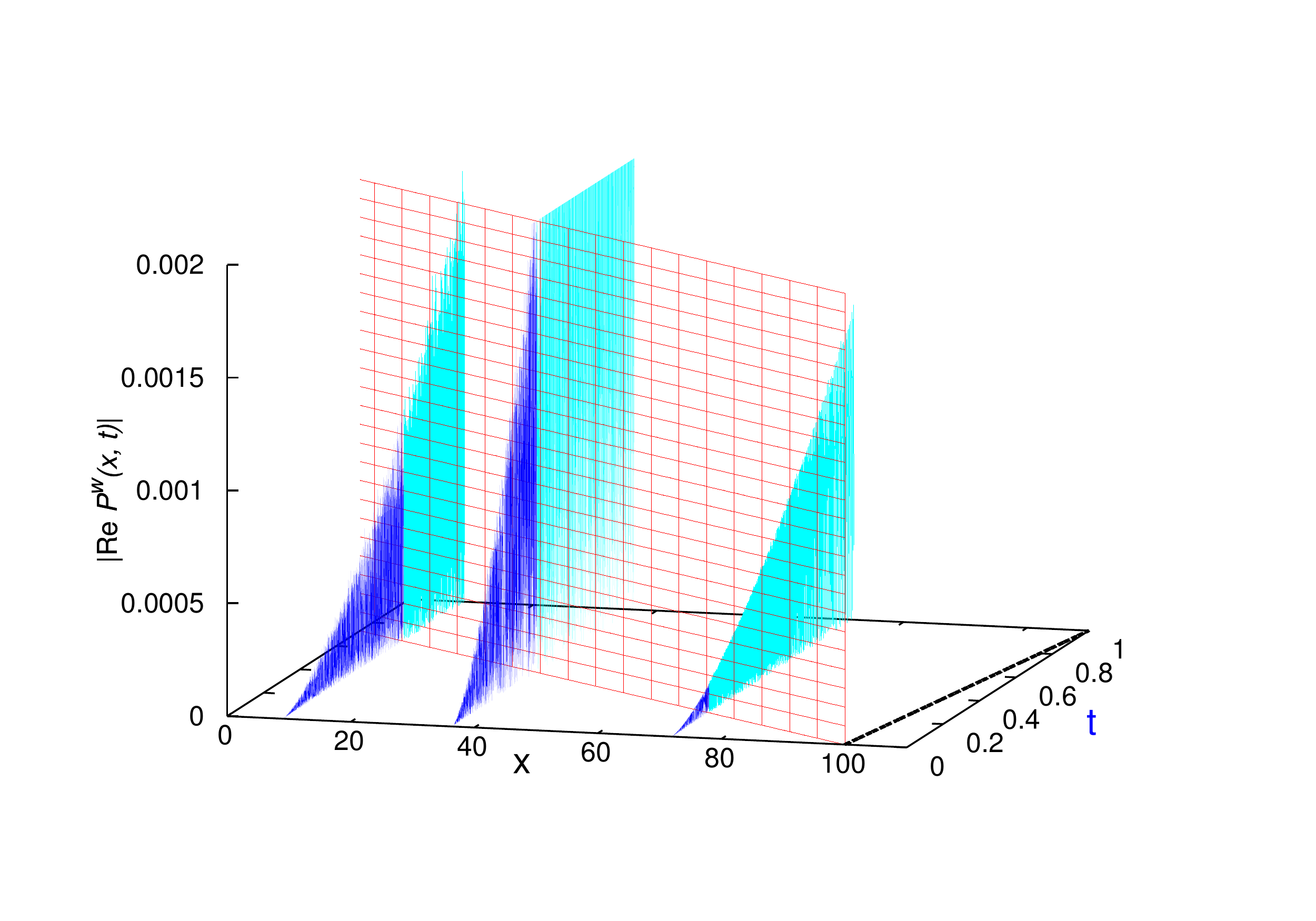} \caption{Time evolution of
the weak value of the momentum of an electron in an expanding cavity of initial length
$L_{0}=100$ $au$. The blue curves represent $|\operatorname{Re}{P^{w}}(x,t)|$
(where $P^{w}$ is the momentum weak value, see Eq. (\ref{pw})) obtained by
making a weak measurement at the corresponding value of $x $. The wall is
initially at $L_{0}$ and the red-gridded plane represents the boundary of the
light cone originating from $x=L_{0}$ at $t=0$. Initially $\operatorname{Re}%
{P}^{w}(x,t=0)=0$ everywhere (and would stay as such for a static wall) but is
seen to oscillate before the light cone reaches the points were $P^{w}$ is
determined (the light blue curves represent $P^{w}$ for times inside the light
cone). The dashed line to the right schematically represents the walls motion
given by $L(t)$. The following parameters have been used: $q=0.5$, $m=1$,
initial wavefunction chosen to be an eigenstate of the static well $\phi
_{n}(x,0)$ [see Eq. (\ref{eig})] with $n=11$ (numbers given in atomic units
(au)).}%
\label{fig1}%
\end{figure}

A couple of illustrations are provided in Figs. \ref{fig1} and \ref{fig2},
where $\operatorname{Re}{P}^{w}$ is shown as a function of time. Fig.
\ref{fig1} shows the case of a moving wall when the system is initially
prepared in a given eigenstate $\phi_{n}$ of the cavity at $t=0$. Fig
\ref{fig2} shows instead the evolution of $\operatorname{Re}{P}^{w}$ in a
static cavity when the system is initially prepared in a basis state $\psi
_{n}(x,t=0)$ [Eq. (\ref{fund})].\ In the former case as noted above we will
have a nonzero current density (whereas $j(x,t)=0$ for any $t$ if the walls
had remained fixed). We see indeed in Fig. \ref{fig1} that $\operatorname{Re}%
{P}^{w}$ changes before a light signal reaches the point where the weak
measurement takes place; the light cone boundary $t_{c}=\left(  L_{0}%
-x\right)  /c$ is indicated by the red-gridded plane. This is the signature of
a form of nonlocality induced by the walls' motion.

In the latter case Eqs. (\ref{jfundn}) and (\ref{pw}) imply that if the
initial state is $\psi_{n}(x,t=0)$, then in a moving cavity the weak value
should evolve following%
\begin{equation}
\operatorname{Re}P^{w}(x,t)=\frac{mqx}{L(t)}. \label{wvbasis}%
\end{equation}
This is represented in Fig. \ref{fig2} by the solid black line. In a fixed
cavity instead $\operatorname{Re}{P}^{w}$ will wildly oscillate, as shown by
the blue curves in Fig. \ref{fig2}. Here again the behavior of a distant wall
(remaining static or in motion) affects the weak value of the momentum
instantaneously, i.e. before a light signal emanating from the wall reaches the
point where the weak measurement is made (the light cone boundary appears as
the vertical dotted line in Fig. \ref{fig2}).

\subsection{Weak Measurement Protocol \label{protocol}}

\begin{figure}[ptb]
\includegraphics*[width=8 cm]{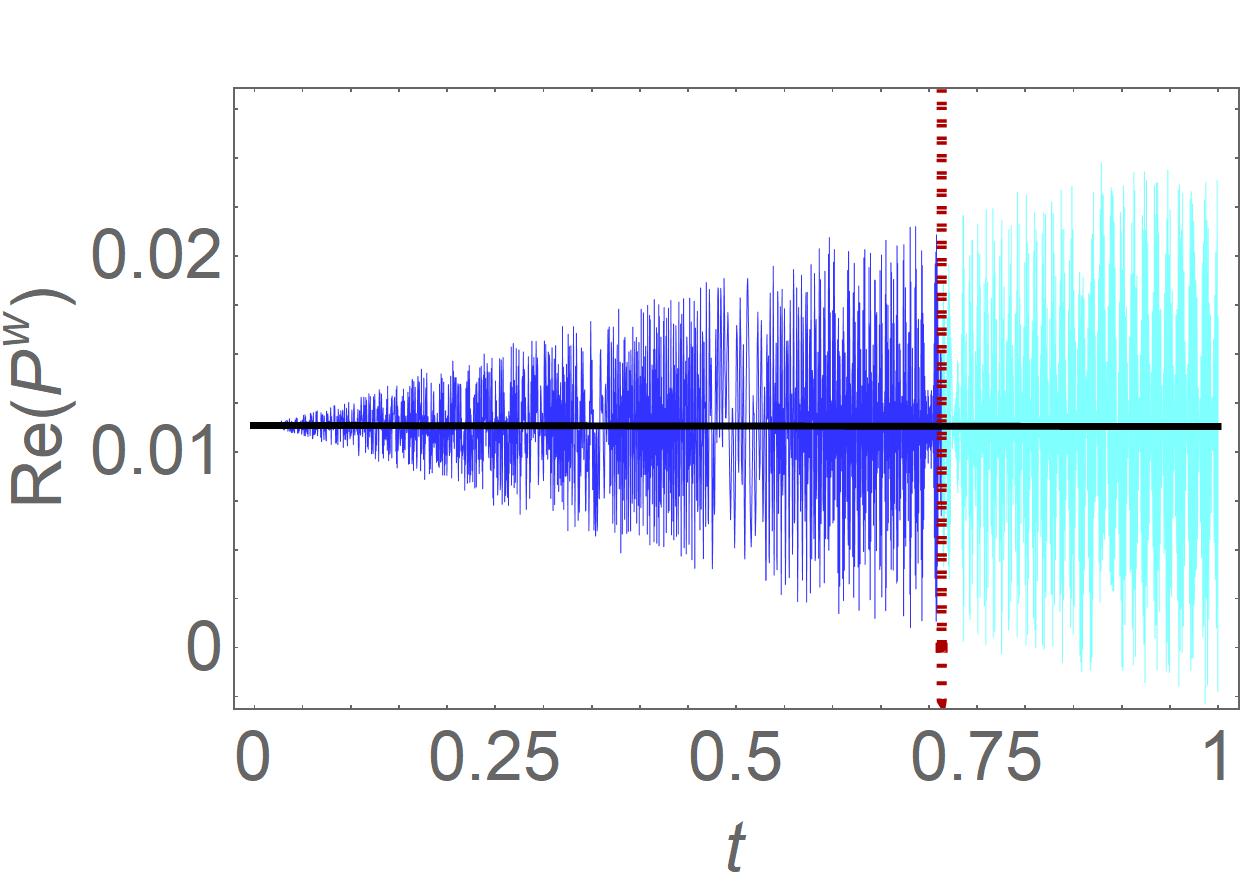} \caption{Time evolution of the
weak value $\operatorname{Re}{P}^{w}(x_{f},t)$ of the momentum for $x_{f}$
lying near the origin (i) in an expanding cavity (solid black line) (ii) in a
static cavity (blue and light blue lines). The red-dashed line is the border
of the light cone (the values of $\operatorname{Re}{P}^{w}(x_{f},t)$ inside
the light cone are in light blue). The fact that the wall moves or remains
fixed is instantaneously reflected in the behavior of the weak value. The
parameters used are $L_{0}=100$, $x_{f}=2.27$, $m=1$, initial wavefunction
chosen to be a basis function $\psi_{n}(x,0)$ [see Eq. (\ref{fund})] with
$n=44$ and in the moving case $q=0.5$ (numbers given in atomic units (au)).}%
\label{fig2}%
\end{figure}

We now introduce the type of protocol that could in principle lead to the
measurement of $\operatorname{Re}{P}^{w}$. An infinite square well is realized
by a long and narrow cavity. The particle (say an electron) is prepared in an
initial state (a given eigenstate of the cavity at $t=0$ or a given basis
function, as in Figs. \ref{fig1} and \ref{fig2} respectively).

At time $t=0$, the sender Alice is located at $x=L_{0}$ and chooses whether to
set the wall in motion (indicating a message bit 1), or to leave the wall at
rest (indicating message bit 0). Then, a very short time later at $t_{1}$,
Bob, located on the opposite side of the cavity near the wall at $x=0$
performs a weak measurement of the position of the particle at location
$x=x_{w}$, followed by a strong measurement of the position at $x_{f}$ (in the
immediate vicinity of $x_{w}$) of the particle at $t_{f}$, again a very short
time later. This procedure is equivalent to measuring the weak value of the
momentum.\ Indeed, it can be shown (see Appendix) that with the weak value of
the position given by%
\begin{equation}
X^{w}=\frac{\left\langle x_{f}\right\vert U(t_{f},t_{w})X\left\vert \psi
(t_{w})\right\rangle }{\left\langle x_{f}\right\vert \left.  \psi
(t_{f})\right\rangle }\label{avn}%
\end{equation}
where $U(t_{f},t_{w})$ is the evolution operator between $t_{w}$ and $t_{f},$
we have%
\begin{equation}
P^{w}=\lim_{t_{f}\rightarrow t_{w}}\frac{m}{t_{f}-t_{w}}\left(  x_{f}%
-X^{w}\right)  .\label{avp}%
\end{equation}
Note that the protocol can also be implemented with a direct weak measurement of the momentum, rather than the two position measurements leading to Eq. (\ref{avp}). This type of weak measurement of the momentum relies on a particular coupling between the particle momentum and an external pointer (in practice, the external pointer is often another degree of freedom of the particle). The important point is that Bob carries out the weak measurement procedure
before a light signal sent by Alice at $t=0$ reaches him, that is we must have
$t_{f}<L_{0}/c$.

Bob can measure whether Alice sent a bit 0 or a bit 1.\ When the initial state
is a stationary state $\phi_{n}(x,0)$ of the cavity $P^{w}=0$ if the wall
remains fixed, but takes a nonzero value (as in Fig.\ \ref{fig1}) if the wall
was set in motion. When the intial state is a basis state $\psi_{n}(x,0)$,
$\operatorname{Re}{P}^{w}$ is given by Eq. (\ref{wvbasis}) if the wall moves,
so Bob can verify by making successive weak measurements on the system if
$\operatorname{Re}{P}^{w}$ is consistent or departs from Eq. (\ref{wvbasis}),
as displayed in Fig.\ \ref{fig2}.

Nevertheless, weak measurements are noisy, and it is impossible for Bob to
learn Alice's choice of signal bit in a single run of the experiment. First
the postselection probability $\int_{x_{w}-\epsilon}^{x_{w}+\epsilon}%
|\psi(x,t_{f})|^{2}dx$ is very small ($\epsilon$ is the width over which the
weak measurement takes place). Second, by definition, when weakly measuring
observable $X$, the pointer wavefunction incurs small shifts [see Eq.
(\ref{Hint})] relative to its width, so many runs of the same experiment will
be necessary in order to extract the weak values. Third even in a weak measurement there is inevitably a back action of the coupling interaction on the subsequent evolution leading to the post-selection. The universal part of the back action is encoded in the imaginary part of the weak value \cite{botero}. A large imaginary part will distort the external pointer state and make even more difficult to extract the small shift of the pointer wavefunction. We note here that for an initial state of the form given by Eq. (\ref{fund}), for which we computed $\operatorname{Re}P^{w}(x,t)=mqx/L(t)$ [Eq. (\ref{wvbasis})], we have

\begin{equation}\label{impw}
\operatorname{Im}P^{w}=-\frac{\hbar \pi n}{L(t)}\cot \left( \frac{n\pi x}{L(t)}\right) .
\end{equation}
The real and imaginary parts of $P^w$ follow a different behavior, and Eq. (\ref{impw}) gives an indication of initial states and spatial regions minimizing the back action. 

However, unlike many other cases where quantum uncertainty prevents
superluminal signaling, we will see that this setup has no such limitation.
This is a serious problem that calls for a more detailed discussion of
non-locality and no-signaling, and compels us to explore possible artifacts
of the model.



\section{Discussion}

\label{discussion}

\subsection{Relativity and non-locality}

The difficulties in reconciling quantum mechanics and special relativity are
well-known \cite{peres RMP}.\ These difficulties stem from the global
character of the state vector, defined in a mathematical configuration space
and not in physical space. It is generally accepted that quantum correlations
cannot lead to superluminal communication of information (no-signaling) as
this would indeed result in an open conflict with relativity. Instead, a
``peaceful coexistence'' \cite{shimony} between quantum mechanics and
relativistic constraints is advocated: as long as one does not attempt to
understand how the quantum correlations come about (in particular through a
hidden-variable model or by endowing the state vector with physical reality),
the observed statistics predicted by quantum theory respect no-signaling.
However if the state vector is assumed to be linked to a real process, then
individual events are difficult to reconcile with relativistic invariance.
This is the case for the collapse of the state vector upon measurement
\cite{aharonov-albert}, or for sub-quantum theories such as the Bohmian model
(see Sec. \ref{bohm} below).

The apparent non-locality seen in the infinite well with a moving wall
investigated here conflicts with this view. The reason is that the non-local
effect comes about as the direct result of a change in a single-particle state
vector (rather than a multi-particle entangled state).\ We will first modify
the weak measurement protocol given above in Sec. \ref{protocol} to show how
it can lead to signaling. We will then discuss the possible artifacts that
could explain our results.

\begin{figure}[ptb]
\includegraphics[angle=0,origin=c,width=16cm]{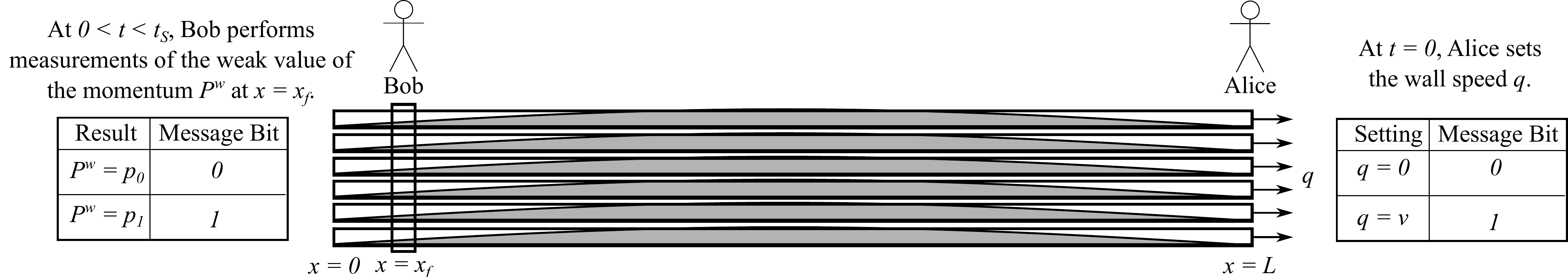}
\caption{A \emph{gedanken} experiment enabling faster-than-light communication
using a large ensemble of long narrow cavities, each containing a single
particle prepared in the same state $\Psi$ (none of the particles are
entangled). Alice sends her message at $t=0$ and Bob receives the message at
$0<t<t_{S}$, where $t_{S}$ is the time for a light signal sent by Alice at
$t=0$ to reach Bob, thus violating the no-signaling principle.}%
\label{fig3}%
\end{figure}

\subsection{Bypassing no-signaling with weak measurements\label{modprot}}

Again, the noisy nature of the weak measurements in the protocol of Sec.
\ref{protocol} does not enable discrimination between the cases of a moving
wall and fixed wall in a single run. However, since only Bob makes the weak
measurement and the strong postselection measurement, we can couch the
statistical argument involving many runs in terms of a single experiment
involving many copies of the system, as shown in Fig. \ref{fig3}.

Consider a \emph{gedanken} experiment taking the form of a very large ensemble
of extremely long and narrow cavities, all aligned together, each with a
particle prepared in the same suitably-chosen state $\Psi$ of the cavity. At
time $t=0$, the sender Alice, located at $x=L_{0}$ chooses either to set all
of those walls in motion (indicating a message bit 1), or to leave them all at
rest (indicating message bit 0). Then, as in Sec. \ref{protocol}, Bob, located
on the opposite side of the cavity near the wall at $x=0$ performs on each
cavity a weak measurement of the position of the particle at location
$x=x_{w}$, followed a very short time later by a strong measurement of the
position at $x_{f}$ of the particle at $t_{f}$.

Next, Bob considers the weak measurement data for the sub-ensemble of cavities
wherein postselection (detection of the particle at $x_{f}$) was successful.
The data allows him to infer the real part of the weak value $P^{w}$ at
$x_{w}$. If the walls are moving, Bob detects the weak momentum $p_{1}$,
indicating a message bit of 1, and otherwise Bob detects the weak momentum
$p_{0}$, indicating a message bit of 0. The values of $p_{0}$ and $p_{1}$
depend on the initial state $\Psi$ in the cavities, which is ideally chosen to
make them easily distinguishable, as with $\phi_{n}$ (Fig. \ref{fig1}) and
$\psi_{n}$ (Fig. \ref{fig2}).

In principle we can make the ensemble arbitrarily large, and the cavities
arbitrarily long, allowing Bob to have enough time to collect sufficient weak
measurement data before a light signal sent from Alice at $L_{0}$ reaches him.
Hence this device enables Alice to send a signal to Bob faster than the speed
of light. All of this analysis is built upon the basis solutions $\psi
_{n}(x,t)$, and thus it appears that these solutions must be nonphysical if
local causality is to be respected. In the following sections we discuss
possible reasons that these solutions are flawed.

\subsection{Sources of nonlocality and possible artifacts}

The most obvious candidates to account for the apparent nonlocality examined
in this work would be the action at a distance effects allowed by a
nonrelativistic formalism such as the Schr\"{o}dinger equation. There are two
types of sources that give rise to superluminal features.\ First, the
existence of instantaneous potentials.\ Second, the fact that a
nonrelativistic framework does not place any restriction on the energy (and
hence velocity) components of a given wavefunction. We argue why these two
features can be discarded in accounting for the results obtained here. We then
examine other possible artifacts, including the fact that unlike a wave
equation of the d'Alembert type, the Schr\"{o}dinger equation does not impose
a particular wave speed, which enables waves to propagate instantaneously.

\subsubsection{Instantaneous potentials}

The instantaneous propagation of potentials appears at first as irrelevant to
our problem: inside the box (except in the vicinity of $x=L(t)$), and in
particular in the region close to $x_{f}$, the potential remains zero at all
times. The potential is therefore not modified so that the question of its
instantaneous propagation appears to be moot.

Nevertheless it should be mentioned that formally, the proper way of obtaining
the basis solutions (see \cite{A2018} and Refs. therein)\ given by Eq.
(\ref{fund}) involves a time-dependent unitary transformation mapping the
moving boundaries problem to a different system with fixed boundaries. The
Hamiltonian of the transformed system is%
\begin{equation}
h(t)=\frac{P^{2}L_{0}^{2}}{2mL^{2}(t)}-\frac{\partial_{t}L(t)}{2L(t)}\left(
XP+PX\right)
\end{equation}
with vanishing boundary conditions at both ends of the interval $[0,L_{0}]$.
$h(t)$ can be understood as describing a system in a fixed wall infinite
potential well with a time-dependent mass and subjected to a time and velocity
dependent potential. In this mapped system -- that, contrary to the original
problem, is described in a single well-defined Hilbert space -- the
instantaneous and uniform character of the time-dependence is obvious. It is
however unlikely that one can make valid inferences concerning the physics of
the original system from the physics of the mapped system (for example the
issue of signaling does not even arise in $h(t)$).\ Quite the contrary, the
global and time-dependent aspects of the mapped system are readily understood
as unphysical features due to the dilation imposed by the unitary
transformation on the original system.

\subsubsection{Infinite velocities}

The issue of infinite velocities arises because in the nonrelativistic
framework a given wavefunction may contain, when expanded over the energy
eigenstates, high energy states that can account for faster than light
propagation. This artifact, due to the nonrelativistic nature of the
Schr\"{o}dinger equation, has been known to produce apparent superluminal
propagation in several instances, in particular when the wavefunction has a
discontinuous cut-off.\ For example in the quantum shutter problem
\cite{moshinsky}, it was shown \cite{garcia-calderon} that a superluminal
propagation occurs due to the high frequencies needed to account for the
cutoff of the initial wavefunction, before the shutter is released.

In the present problem, the contribution of high energy states can be
evaluated by expanding the solution of the Schr\"{o}dinger equation
$\psi(x,t)$ as given by Eq. (\ref{expan}) in the instantaneous eigenstate
basis. Let us assume that a cavity is initially in the fixed wall eigenstate
$\phi_{n_{0}}(x,t=0)$ [cf. Eq. (\ref{eig})]. In order to compute the
evolution, we need to expand $\phi_{n_{0}}(x,t=0)$ over the basis functions
$\psi_{k}(x,t=0),$ as in Eq. (\ref{expan}) but here with a single term $n_{0}%
$. The overlap coefficients%
\begin{equation}
d_{kn}(t)=\int_{0}^{L(t)}\psi_{k}^{\ast}(x,t)\phi_{n}(x,t)dx
\end{equation}
can be readily computed in closed form as%
\begin{equation}
d_{kn}(t)=\frac{e^{-i\pi/4}\sqrt{\hbar\pi}}{2\sqrt{2M}}\left(
\begin{array}
[c]{c}%
e^{\frac{i\pi^{2}\hbar\left(  k^{2}qt+L_{0}(k-n)^{2}\right)  }{2L_{0}M)}%
}\left[  \operatorname{erf}\left(  \frac{e^{i\pi/4}(\pi\hbar(k-n)+M)}%
{\sqrt{2\hbar M}}\right)  +\operatorname{erf}\left(  \frac{e^{i\pi/4}(\pi
\hbar(n-k)+M)}{\sqrt{2\hbar M}}\right)  \right] \\
-e^{\frac{i\pi^{2}\hbar\left(  k^{2}qt+L_{0}(k+n)^{2}\right)  }{2L_{0}M)}%
}\left[  \operatorname{erf}\left(  \frac{e^{i\pi/4}(\pi\hbar(k+n)+M)}%
{\sqrt{2\hbar M}}\right)  -\operatorname{erf}\left(  \frac{e^{i\pi/4}(\pi
\hbar(k+n)-M)}{\sqrt{2\hbar M}}\right)  \right]
\end{array}
\right)  \label{dcoef}%
\end{equation}
with $M\equiv mqL(t).$

We first need to determine $d_{kn_{0}}(t=0).$ Assuming $n_{0}$ is small, each
square bracket in Eq. (\ref{dcoef}) is seen to vanish for $k\gg M/\pi\hbar$
when the error functions cancel out, so the infinite sum in Eq. (\ref{expan})
can actually be cut off at a value somewhat larger (depending on the mass)
than $\bar{k}=M/\pi\hbar$. The contribution of an instantaneous eigenstate
$\phi_{n}(x,t)$ at time $t$ is then given by%
\begin{equation}
\left\langle \phi_{n}(t)\right\vert \left.  \psi(t)\right\rangle =\sum
_{k=1}^{k_{\text{cut}}}d_{kn_{0}}(0)d_{kn}^{\ast}(t) \label{projs}%
\end{equation}
where $k_{\text{cut}}$ is the cutoff value. From Eq. (\ref{dcoef}) it is seen
that the error functions in the brackets will cancel each other for $n\gg
\bar{k}$. The modulus of the velocity in an instantaneous eigenstate $\phi
_{n}(x,t)$ is given by
\begin{equation}
v(n,t)=\hbar\pi n/mL(t) \label{velo}%
\end{equation}
and for $n=\bar{k}$, this becomes $v(\bar{k})$ $=q$. Therefore, depending on
the mass, we can expect from the properties of the error function that the
highest energy eigenstates that will contribute will have at most a
corresponding velocity 2 or 3 orders of magnitude larger than $q$, the
velocity at which the wall is expanding, which can be arbitrarily smaller than
$c$.

Mathematically the tail remains as $d_{kn}$ falls off as $1/n^{3}$ for large
$n$ and is not strictly zero.\ The physical effects relevant to the tail can
generally be ignored, at least as quantities related directly to the
wavefunction are concerned (for instance when comparing the wavefunction
evolution $\psi(x,t)-\psi(x,0)$). This can be confirmed numerically. Let us
consider an electron in a cavity with the parameters given in Fig.\ \ref{fig1}%
. To quantify the tails, we note that
\begin{equation}
\sum_{k=1}^{k_{\text{cut}}}\left\vert d_{kn_{0}=11}(0)\right\vert ^{2}=1
\end{equation}
holds with $k_{\text{cut}}=650$ if the numerical zero is set at $10^{-10}$
(meaning that the states with $k>650$ account for a relative part of less than
$10^{-10}$ in the total state vector). At some arbitrary time $t$, the
solution $\psi(x,t)$ is expanded over the instantaneous eigenstates
(\ref{eig}). Now
\begin{equation}
\sum_{n=1}^{n_{\text{cut}}}\left\vert \sum_{k=1}^{k_{\text{cut}}}
d_{kn_{0}=11}(0) d_{nk}^{*}(t) \right\vert ^{2}=1
\end{equation}
holds again up to $10^{-10}$ for $n_{\text{cut}}=650$ for, say $t=t_{S}/2$ (a
light signal sent from the right end of the well is halfway through in the
fixed wall frame).\ According to Eq. (\ref{velo}) $v(n=650,t_{S})/c<15\%$.
Hence if we neglect the tail, we recover the wavefunction up to one part in
$10^{-10}$ while remaining  far from the regime of superluminal velocities.

However concerning the current density, the quantities we are looking at are
very small and convergence to the same numerical zero taken for the
wavefunction is achieved only when including $d_{kn}(t)$ coefficient lying in
the tail. For example when the initial state is $\phi_{n_{0}}(x,t=0)$, the
current density in the moving cavity is initially zero, and for small $x$
values $j(x,t)$ rises slowly. The determination of $j(x,t)$, which needs to be
done by expanding over the basis functions, converges by including expansion
coefficients lying in the tail and corresponding to velocities with
arbitrarily high energies. To be clear, $j(x,t)$ is nonzero if coefficients in
the tail are excluded, but convergence is only achieved when coefficients in
the tail are included. In this sense, going into the tail (i.e., including
arbitrarily high energies into the computation) appears as a mathematical
requirement to obtain a stable result rather than giving rise to the
phenomenon itself.

We can therefore conclude that the nonlocal effect put into evidence above
does not appear to be due to the existence of arbitrarily high velocities that
would propagate the change in the quantum state at an arbitrarily high
velocity. The ambiguity in the inclusion of the tail terms (necessary to
achieve convergence for the initial states we have worked with here) can only
be lifted by working within a fully relativistic framework.

\subsubsection{Model artifact and wave equations\label{artefact}}

Mathematically, the source of nonlocality is straightforward to pinpoint.\ As
can be read off from Eqs. (\ref{pw}) or (\ref{avp}), the instantaneous change
of the wavefunction at every point of space as time unfolds (due to the
expansion of the cavity) is what changes the current density or the momentum
weak value. From a formal point of view, the peculiarity of the model, as
mentioned in Sec. \ref{sec2}, is that at each time $t$, the system is defined
on a different Hilbert space. Following the system evolution as time unfolds
implies connecting vectors belonging to different Hilbert spaces. Hence,
connecting \textquotedblleft independent\textquotedblright\ solutions
belonging to different Hilbert spaces might result in an unphysical picture,
resulting in fictitious instantaneous effects. Note that such a phenomenon is
expected to be ubiquitous when the potential changes in a specific restricted
region of space: the state vector changes instantaneously in Hilbert space,
leading to a modification of the wavefunction in the regions in which the
potential was not modified.

This is related to the fact that from the point of view of wave equations, we
know that the Schr\"{o}dinger equation does not impose a finite propagation
velocity. It might be conjectured that by supplementing the Schr\"{o}dinger
equation with a propagation velocity, as is the case of the relativistic
Klein-Gordon equation or d'Alembert equation for Maxwell fields, we would get
rid of these nonlocal effects. We can pursue the analogy with classical
electromagnetism further: from the point of view of Maxwell fields in an
expanding cavity, we would see the instantaneous change of standing waves as
an artifact of the model. Indeed, we can rely on Maxwell's equations to take
into account the transient effects (radiation and propagation) due to the
moving charges composing the wall. However in standard unitary quantum
mechanics there are no fundamental equations underlying the Schr\"{o}dinger
evolution on which we could rely to take into account this specific transient
effect. Note that the basis functions (\ref{fund}) are exact solutions of the
Schr\"{o}dinger equation: they play the same role in the present problem as
the Moshinsky function in the paradigmatic shutter problem
\cite{moshinsky,muga}.\ The Moshinsky functions form the transient basis
solutions of the Schr\"{o}dinger equation in the shutter problem \cite{muga}.
It thus looks like taking into account a putative transient phenomenon for our
moving wall problem would need to supplement the standard quantum formalism.

\subsection{Signaling and the Bohmian model\label{bohm}}

As we noted below Eq. (\ref{pw}), the real part of the momentum weak value
$P^{w}$ is essentially the velocity of the particle postulated in the de
Broglie-Bohm interpretation \cite{holland} (in short dBB or Bohmian model).
Recall that dBB accounts for quantum phenomena by postulating the existence of
point-like particles guided by the wavefunction. If we write the wavefunction
in polar form as
\begin{equation}
\psi(x,t)=\rho(x,t)\exp(i\sigma(x,t)/\hbar), \label{5}%
\end{equation}
then $\rho$ and $\sigma$ obey the equation
\begin{equation}
\frac{\partial\sigma}{\partial t}+\frac{(\mathbf{\triangledown}\sigma)^{2}%
}{2m}+V+Q=0 \label{6}%
\end{equation}
where $V$ is the usual potential and the term
\begin{equation}
Q(x,t)\equiv-\frac{\hbar^{2}}{2m}\frac{\partial_{x}^{2}\rho}{\rho} \label{qp}%
\end{equation}
is known as the quantum potential. The particle velocity, defined from within
the Bohmian model by $v(x,t)=\partial_{x}\sigma(x,t)/m$ rather than the
equivalent Eq. (\ref{bv}) obeys a Newton law modified by the presence of the
quantum potential:%
\begin{equation}
m\frac{dv}{dt}=-\partial_{x}(V+Q). \label{e7}%
\end{equation}
Bohmian trajectories in systems analogous to the one investigated here have
been previously computed \cite{mousavi2014}.

The Bohmian model is generally recognized as being nonlocal.\ The culprit is
the quantum potential, whose local value depends on the instantaneous
positions of all particles in the universe. Hence, a Bohmian particle is
instantaneously affected by the motion of all of those particles, including
those which produce the effective barriers of a potential well. Nevertheless
it is generally accepted by proponents of the model that this nonlocality
cannot be used to communicate due to the intrinsic quantum randomness (which
also includes but is not limited to the ignorance of the particle's initial
condition in a given realization). There is therefore nonlocality at the
individual level, but because of the random character of quantum mechanics,
no-signaling holds at the statistical level.

In the Bohmian account of our device introduced in Sec.\ \ref{protocol}, the
postselected particle at $x_{f}$ was assumed to be there even before it was
detected.\ However its dynamics were affected by the quantum potential, which
carries the influence of the far wall's motion. Here, the superluminal
signaling aspect of our protocol yields a conflict with the usual notion that
dBB, despite being explicitly nonlocal, obeys the no-signaling principle. For
instance in the EPR-Bell setting involving two particles in an entangled
state, the quantum potential changes instantaneously, thereby
\textquotedblleft contradicting the spirit of relativity\textquotedblright%
\ (as put nicely by Holland, cf Sec.\ 11.3 of \cite{holland}), although this
change has no observable consequences. In our system, the only quantity that
changes instantaneously is the quantum potential given by\ Eq. (\ref{qp}),
since as we remarked above, the usual potential remains constant except in the
vicinity of the wall, but there are observable consequences. Note that the
enforcement of a finite propagation velocity mentioned in Sec. \ref{artefact}
in dBB would constitute a constraint on the propagation speed of the quantum
potential itself.

\section{Conclusion\label{conclu}}

To summarize, we have put forward a model displaying an apparent
single-particle nonlocality and enabling faster-than-light communication. As
discussed in Sec. \ref{discussion}, we believe these effects are artifacts of
the fact that we are using the nonrelativistic Schr\"{o}dinger equation, and
not genuine physical effects that will one day be realized as actual
communication devices. However, even after this analysis, we still cannot
claim to have pinned down exactly why the Schr\"{o}dinger equation violates
local causality, even in a regime where it seems relativistic effects should
be negligible.

From a physical standpoint it seems likely that there must be a transient
behavior which begins at the moving wall and propagates through the
wavefunction at or below $c$. These transient behaviors are not given by the
solutions of the standard Schr\"{o}dinger equation, so it seems plausible to
suggest that the correct dynamical evolution equation would contain terms
accounting for such transients. Even presuming this is the right way to impose
a relativistic constraint on the non-relativistic Schr\"{o}dinger equation, we
do not presently have a suggested form for incorporating this constraint. A
relativistic treatment would be helpful, despite the well-known limitations
affecting the single particle relativistic wavefunctions.

To conclude, the system investigated in this work raises interesting questions
about the general trustworthiness of any solutions to the Schr\"{o}dinger
equation involving time-dependent potentials localized in a given spatial
region but affecting the entire wavefunction. We remain open to the
possibility that there could be some other explanation that we have not
considered, and we would be very pleased if a more complete resolution of this
conundrum could be found.\newline

\textbf{Acknowledgments:} AM thanks the participants of a FQXI workshop in
Marseille (July 2017) for useful exchanges on single particle nonlocality. MW
acknowledges partial support from the Fetzer Franklin Fund of the John E.
Fetzer Memorial Trust.

\appendix
\section*{Appendix: Weak value of the momentum in terms of position measurements}

We prove here Eq. (\ref{avp}) expressing $P^{w}$ in terms of a weak and then a
projective position measurements. The manipulations are similar to the ones
employed in Ref. \cite{wiseman}. With $\Delta t\equiv t_{f}-t_{w}$ very small,
we have%
\begin{equation}
X^{w}=\frac{\left\langle x_{f}\right\vert U(t_{f},t_{w})X\left\vert \psi
(t_{w})\right\rangle }{\left\langle x_{f}\right\vert \left.  \psi
(t_{f})\right\rangle }=\frac{\left\langle x_{f}\right\vert \left(  1-\frac
{i}{\hbar}\Delta tH\right)  X\left\vert \psi(t_{w})\right\rangle
}{\left\langle x_{f}\right\vert \exp\left(  -\frac{i}{\hbar}\Delta tH\right)
\left\vert \psi(t_{w})\right\rangle }.
\end{equation}
Since $i\left[  H,X\right]  =\hbar P$ we have%
\begin{align}
X^{w}  & =\frac{\left\langle x_{f}\right\vert \left(  X-\frac{\Delta t}%
{m}P-\frac{i\Delta t}{\hbar}XH\right)  \left\vert \psi(t_{w})\right\rangle
}{\left\langle x_{f}\right\vert \exp\left(  -\frac{i}{\hbar}\Delta tH\right)
\left\vert \psi(t_{w})\right\rangle }\\
& =\frac{x_{f}\left\langle x_{f}\right\vert \left.  \psi(t_{w})\right\rangle
-\frac{\Delta t}{m}\left\langle x_{f}\right\vert P\left\vert \psi
(t_{w})\right\rangle -x_{f}\left\langle x_{f}\right\vert \left(  1-\exp\left(
-\frac{i}{\hbar}\Delta tH\right)  \right)  \left\vert \psi(t_{w})\right\rangle
}{\left\langle x_{f}\right\vert \exp\left(  -\frac{i}{\hbar}\Delta tH\right)
\left\vert \psi(t_{w})\right\rangle }\\
& =\frac{-\frac{\Delta t}{m}\left\langle x_{f}\right\vert P\left\vert
\psi(t_{w})\right\rangle }{\left\langle x_{f}\right\vert \exp\left(  -\frac
{i}{\hbar}\Delta tH\right)  \left\vert \psi(t_{w})\right\rangle }+x_{f}.
\end{align}
Therefore%
\begin{equation}
\frac{\left\langle x_{f}\right\vert P\left\vert \psi(t_{w})\right\rangle
}{\left\langle x_{f}\right\vert \exp\left(  -\frac{i}{\hbar}\Delta tH\right)
\left\vert \psi(t_{w})\right\rangle }=\frac{m}{\Delta t}\left(  x_{f}%
-X^{w}\right)
\end{equation}
and taking the limit $\Delta t\rightarrow0$ (or equivalently $t_{f}\rightarrow
t_{w})$ we recover Eq. (\ref{avp}).

\end{document}